\documentclass[aps,prd,tightenlines,preprint,nofootinbib,a4paper]{revtex4}
\usepackage{epsfig}

\begin{document}

\title{Features of Neutrino Mixing}

\author{S. H. Chiu\footnote{schiu@mail.cgu.edu.tw}}
\affiliation{Physics Group, CGE, Chang Gung University, 
Taoyuan 33302, Taiwan}

\author{T. K. Kuo\footnote{tkkuo@purdue.edu}}
\affiliation{Department of Physics, Purdue University, West Lafayette, IN 47907, USA}

\begin{abstract}
The elements (squared) of the neutrino mixing matrix are found to satisfy, as functions of the induced mass, 
a set of differential equations. They show clearly the dominance of pole terms when the neutrino masses
``cross".  Using the known vacuum mixing parameters as initial conditions, it is found that these equations
have very good approximate solutions, for all values of the induced mass.  The results are applicable to
long baseline experiments (LBL).

\end{abstract}


\maketitle

\pagenumbering{arabic}



\section{Introduction}

Advances in neutrino oscillation experiments have yielded a wealth of information on the intrinsic neutrino properties,
their masses and mixings. Two mass differences are well measured.  Neutrino mixing is described by the $ 3\times 3$
unitary PMNS matrix, $V_{PMNS}$, which, because of rephasing invariance, contains only four
physical variables.  Thus, instead of the matrix elements $V_{\alpha i}$ ($\alpha=e,\mu,\tau$; $i=1,2,3$), 
only rephasing invariant combinations thereof, such as $|V_{\alpha i}|$ or $|V_{\alpha i}|^{2}=W_{\alpha i}$, are
physically measurable, and they can be expressed in terms of four physical parameters.
In an ideal situation, where all the parameters are precisely known, one can choose to use any set and arrive at the same result for the exact $|V_{\alpha i}|$ values.  In reality, however, our knowledge about $|V_{\alpha i}|$ is far from
uniform. While the elements $|V_{ei}|$, ($i=1,2,3$), and $|V_{\alpha 3}|$, ($\alpha=\mu,\tau$) (group I), are
experimentally accessible and well determined, the remaining four elements 
($|V_{\mu1}|,|V_{\mu2}|,|V_{\tau 1}|,|V_{\tau 2}|$) (group II) suffer from large uncertainties 
(see, e.g. \cite{Forero:2014bxa,Gonzalez-Garcia:2015qrr,Capozzi:2016rtj}).  
In the widely used standard parametrization (SP), the elements in group I are simple functions of SP,
so that the angles ($\theta_{12},\theta_{23},\theta_{13}$) are all well determined,
while elements in group II are complicated functions of SP, making it very hard to estimate the remaining phase 
(the Dirac $\delta$).  Note that the unitarity conditions on $W_{\alpha i}=|V_{\alpha i}|^{2}$, which may help to constraint
group II elements, are hard to implement in terms of SP.   The above comments are    
given assuming neutrinos are Dirac particles.
However, there are likely scenarios in which neutrinos are Majorana particles,
and there may be extra neutrinos in addition to those in the standard model \cite{Schechter:1980gk}.
In the first case, phases of Majorana neutrinos are physical, and rephasing invariance is lost.
In the second case, unitarity of $W_{\nu}$ is broken.  Thus, the results in this paper are valid
only if extra neutrinos do not exist, and, for Majorana neutrinos, only when we consider neutrino oscillations which
conserve the lepton number.

In this paper, we propose to parametrize $W$ directly and simply.  The linear dependence of these parameters facilitates the implementation of the unitarity condition so that the group II elements are woven into the structure of $W$.
It is found that, given the known values of group I, these elements are already significantly constrained. 
They are also tightly correlated. For neutrino propagation in matter, we establish a set of differential equations
for the evolution of the elements $W_{\alpha i}$, as functions of the induced neutrino mass.
These equations are simple and compact in form, so that one can visualize the properties of their solutions with ease.
It is found that the result corresponds to two well-separated level-crossing solutions.  The mixing parameters change
rapidly only in the neighborhood of two resonances, while in the regions outside of those they are mostly flat.
Another interesting  consequence of level crossing is the decoupling effects, which tend to suppress the 
influence of initial conditions.  Thus, it will be shown that the mixing matrix in matter is actually simpler
than that in vacuum.   


This paper is organized as follows.
In Section II, the general properties of a set of rephasing-invariant parameters are briefly introduced.
In Section III, the physical variables $|V_{\alpha i}|^{2}=W_{\alpha i}$ are parametrized by imposing the unitary
conditions.  In Section IV, a set of differential equations for matter effects are derived, and the approximate solutions are obtained. The numerical solutions for the differential equations are shown in Section V.  We then outline the possible applications of our formulations to experiments in Section VI and summarize this work in Section VII. 

\section{Notations}

It is well known that physical observables are independent of rephasing
transformations on the mixing matrices of quantum-mechanical states.
Whereas there is nothing wrong with using these matrix elements in intermediate steps 
of a calculation, at the end of the day, they must form rephasing-invariant
combinations in physical quantities.
This situation is similar to that in gauge theory, where one often resorts
to a particular gauge choice for certain problems.
The final results, however, must be gauge invariant.  In this paper,
we propose to use, from the outset, parameters that are rephasing invariant.  
The use of only physical variables has another interesting consequence.
As we will show in Sec. IV, as functions of the induced neutrino mass, the physical variables
obey a simple set of differential equations, while one expects that they would be rather complicated
when written in terms of the SP variables.
Note that there is a similar simplification for the RGE of neutrinos and 
of quarks \cite{Chiu:2015ega,Chiu:2016qra}.

We turn now to Ref. \cite{Kuo:2005pf}, where it was pointed out that six rephasing invariant combination 
can be constructed from a $3 \times 3$ unitary mixing matrix $V$, which, for $\nu$ mixing, are given by
\begin{equation}
\Gamma_{ijk}=V_{\alpha i}V_{\beta j}V_{\gamma k}=R_{ijk}-iJ,
\end{equation}
where $(\alpha,\beta,\gamma)=(e,\mu,\tau)$, 
($i,j,k$) are cyclic permutations of $(1,2,3)$, and $detV=+1$ is imposed.
The common imaginary part is identified with the Jarlskog invariant \cite{Jarlskog:1985ht}, 
and the real parts are defined as
\begin{equation}\label{Rijk}
(R_{123},R_{231},R_{312};R_{132},R_{213},R_{321})=(x_{1},x_{2},x_{3};y_{1},y_{2},y_{3}).
\end{equation}
The $(x_{i},y_{j})$ parameters are bounded, $-1\leq (x_{i},y_{j}) \leq 1$,
with $x_{i} \geq y_{j}$ for any pair of $(i,j)$.
It is also found that the six parameters satisfy two conditions,
\begin{equation}\label{con1}
det V=(x_{1}+x_{2}+x_{3})-(y_{1}+y_{2}+y_{3})=1,
\end{equation}
\begin{equation}\label{con2}
(x_{1}x_{2}+x_{2}x_{3}+x_{3}x_{1})-(y_{1}y_{2}+y_{2}y_{3}+y_{3}y_{1})=0,
\end{equation}
leaving four independent parameters for the mixing matrix.
They are related to the Jarlskog invariant,
\begin{equation}
J^{2}=x_{1}x_{2}x_{3}-y_{1}y_{2}y_{3},
\end{equation}
and the squared elements of $V$,
\begin{equation}\label{WV2}
W=[|V_{\alpha i}|^{2}]=
\left(\begin{array}{ccc}
   x_{1}-y_{1} & x_{2}-y_{2} & x_{3}-y_{3} \\
   x_{3}-y_{2} & x_{1}-y_{3} & x_{2}-y_{1} \\
    x_{2}-y_{3} & x_{3}-y_{1} &x_{1}-y_{2} \\
    \end{array}
    \right).
    \end{equation}
The matrix of the cofactors of $W$, denoted as $w$ with $w^{T}W= (\mbox{det}W)I$, is given by
\begin{equation}
w=
\left(\begin{array}{ccc}
   x_{1}+y_{1} & x_{2}+y_{2} & x_{3}+y_{3} \\
   x_{3}+y_{2} & x_{1}+y_{3} & x_{2}+y_{1} \\
    x_{2}+y_{3} & x_{3}+y_{1} &x_{1}+y_{2} \\
    \end{array}
    \right)
    \end{equation}
The elements of $w$ are also bounded, $-1 \leq w_{\alpha i} \leq +1$, and
\begin{equation}
\sum_{i}w_{\alpha i}=\sum_{\alpha}w_{\alpha i}=\mbox{det} W,
\end{equation}
\begin{equation}
\mbox{det} W=\sum x_{i}^{2}-\sum y_{j}^{2}=\sum x_{i}+\sum y_{j}.
\end{equation}   
The relations between $(x_{i},y_{j})$ and the standard parametrization
can be found in Ref.\cite{Chiu:2011rj}.   

There are other rephasing-invariant combinations that are useful.
One first considers the product of four mixing elements \cite{Jarlskog:1985ht}
\begin{equation}\label{eq:piij}
\Pi_{ij}^{\alpha \beta}=V_{\alpha i}V_{\beta j}V_{\alpha j}^{*}V_{\beta i}^{*},
\end{equation}
which can be reduced to 
\begin{eqnarray}
\Pi_{ij}^{\alpha \beta} & = & |V_{\alpha i}|^{2}|V_{\beta j}|^{2}-
 \sum_{\gamma k}\epsilon _{\alpha \beta \gamma}\epsilon_{ijk}V_{\alpha i}V_{\beta j}V_{\gamma k} \nonumber \\
   & = & |V_{\alpha j}|^{2}|V_{\beta i}|^{2}+
 \sum_{\gamma k}\epsilon _{\alpha \beta \gamma}\epsilon_{ijk}V_{\alpha j}^{*}V_{\beta i}^{*}V_{\gamma k}^{*}.
 \end{eqnarray}
In addition, for $\alpha \neq \beta \neq \gamma$ and
$i\neq j \neq k$, we define
\begin{equation}\label{piab}
\Pi^{\alpha \beta}_{ij}\equiv \Pi_{\gamma k}=\Lambda_{\gamma k}+iJ.
\end{equation}
Since $Re(\Pi^{\alpha \beta}_{ij})$ takes the forms,
\begin{equation}
Re(\Pi^{\alpha \beta}_{ij})=|V_{\alpha i}|^{2}|V_{\beta j}|^{2}-x_{a}=
|V_{\beta i}|^{2}|V_{\alpha j}|^{2}+y_{l},
\end{equation}
we have
\begin{equation}\label{lambda}
\Lambda_{\gamma k}=
\frac{1}{2}(|V_{\alpha i}|^{2}|V_{\beta j}|^{2}+|V_{\alpha j}|^{2}|V_{\beta i}|^{2}-|V_{\gamma k}|^{2}).
\end{equation}
In terms of the $(x,y)$ variables,
\begin{equation}\label{lambdaxy}
\Lambda_{\gamma k}=x_{a}y_{l}+x_{b}x_{c}-y_{l}(y_{m}+y_{n}),
\end{equation}
where $(x_{a},y_{l})$ comes from $|V_{\gamma k}|^{2}=x_{a}-y_{l}$, 
and $a\neq b\neq c$, $l\neq m \neq n$.

Another interesting combination is given by 
\begin{equation}
\Xi_{\alpha i}=V_{\alpha j}V_{\alpha k}V_{\alpha i}^{*}V_{\beta i}V_{\gamma i}
=(y_{m}y_{n}-x_{b}x_{c})+iJ(1-|V_{\alpha i}|^{2}).
\end{equation}
Here if $|V_{\alpha i}|^{2}=x_{a}-y_{l}$, then $b\neq c\neq a$, $m \neq n \neq l$.
This means that if one takes the $\alpha$th row and the $i$th column, complex conjugates
the vertex ($V_{\alpha i}^{*}$), then the product is rephasing invariant and has a well-defined imaginary part.
Of particular interest is $\Xi_{e3}$.  If we write $\Xi_{e3}=|\Xi_{e3}|e^{i\delta}$, then
$Im[\Xi_{e3}]=J(1-W_{e3})=\sin \delta |\Xi_{e3}|$.  Thus, the (rephasing invariant) phase
of $\Xi_{e3}$ is identified with the Dirac phase in the SP.  Also, using vacuum values,
$|\Xi^{0}_{e3}|^{2} \cong 1.1 \times 10^{-3}$ and $Re[\Xi^{0}_{e3}] \cong w^{0}_{e3}/2$, 
it was found \cite{Chiu:2012uc} that  
\begin{equation}
(w^{0}_{e3}/2)^{2}+(J^{0})^{2} \cong 1.1 \times 10^{-3}.
\end{equation}
As a result, the leptonic CP violation depends crucially on the determination of $w^{0}_{e3}$.

\section{Parametrization of the neutrino mixing matrix}

Neutrino mixing is described by the $3 \times 3$ unitary PMNS matrix, $V_{PMNS}$, or $V_{\nu}$.
Because of the rephasing invariance, only four parameters contained therein are physical.
If these parameters are all precisely known, using different sets will not make much difference.
In reality, the choice of them depends on how best they can be used to incorporate our partial knowledge
gleaned from available experimental data.  The widely used standard parametrization (SP) emphasizes the
matrix elements $V_{ei}$ and $V_{\alpha 3}$ $(i=1,2,3;\alpha =e,\mu,\tau)$,
because (the absolute squares of) these elements have been well measured.  The remaining elements
($V_{\mu 1}, V_{\mu 2}, V_{\tau 1}, V_{\tau 2}$), however, are complicated functions of the SP
and are hard to pin down 
(see, e.g.,  \cite{Forero:2014bxa,Gonzalez-Garcia:2015qrr,Capozzi:2016rtj}), 
given that their possible errors are not related to those of the SP in any simple way.

In this paper we propose a parametrization by concentrating directly on the physical variables
$|V_{\alpha i}|$ or $|V_{\alpha i}|^{2}=W_{\alpha i}$.  By imposing the unitarity conditions uniformly,
it is seen that the errors of all $W_{\alpha i}$ are strictly and simply correlated.
To include the matter effects, these parameters are considered to be functions of the induced neutrino mass
$A=2\sqrt{2}G_{F}n_{e}E$. In the next section, we show that they obey simple differential equations which,
with the known vacuum neutrino parameters as inputs, have good approximate solutions.

A general parametrization of $W_{\nu}$ can be written in the following form
\begin{equation}\label{nuW}
 [W_{\nu}]= [\widetilde{W}_{0}]+b [\mathcal{B}]+c [\mathcal{C}]+d [\mathcal{D}] +e [\mathcal{E}],
\end{equation}
where we choose
\begin{equation}\label{W0}
[\widetilde{W}_{0}]=\left(\begin{array}{ccc}
 2/3,& 1/3, & 0 \\
 1/6, & 1/3, &1/2 \\
    1/6, & 1/3, &1/2 \\
    \end{array}
    \right) ,
\end{equation}
and 
\begin{eqnarray}
[\mathcal{B}]=\left(\begin{array}{ccc}
   1, & -1, & 0 \\
-1/2, & 1/2, & 0 \\
    -1/2, & 1/2,& 0 \\
    \end{array} 
    \right), \hspace{0.5in}  
    [\mathcal{C}]=\left(\begin{array}{ccc}
  -1/2, & -1/2, & 1 \\
1/4,& 1/4, & -1/2 \\
   1/4, &1/4, &-1/2 \\
    \end{array} \nonumber
    \right),    
    \end{eqnarray} 
    \begin{eqnarray}\label{bcde}
    [\mathcal{D}]=\left(\begin{array}{ccc}
   0,& 0, & 0 \\
-1/2,&-1/2, & 1\\
    1/2,& 1/2, & -1 \\
    \end{array}
    \right) , \hspace{0.5in}
    [\mathcal{E}]=\left(\begin{array}{ccc}
  0, & 0, & 0 \\
 1, & -1, & 0 \\
    -1, & 1, &0 \\
    \end{array}
    \right).    
\end{eqnarray}
Here, the parameters ($b,c,d,e$) are functions of $A$, $b(A)$ etc., to be considered in detail in Sec. IV.
Their vacuum values carry the subscripts 0, e.g., $b(A=0)=b_{0}$, etc.  Also, by construction, 
the unitarity conditions are strictly satisfied by $[W_{\nu}]$.

The constant matrices are chosen to take into account the known features of vacuum neutrino mixing.
In particular, the matrix $[\widetilde{W}_{0}]$ is a well-known approximation to the vacuum mixing matrix $[W_{\nu}(0)]$.
Thus, all the vacuum values ($b_{0},c_{0},d_{0},e_{0}$) are small. Indeed, the estimated values \cite{PDG} 
of ($b_{0},c_{0}$) are $b_{0}\cong 0.01$ and $c_{0} \cong 0.02$.  While the sign of $d_{0}$ is ambiguous,
its absolute value is favored to be $d_{0} \cong 0.05$.  
Also, there is a bound on $w^{0}_{e3}\cong-\frac{1}{6}d_{0}+e_{0}$ \cite{Chiu:2012uc}, 
given by $(-\frac{1}{6}d_{0}+e_{0})^{2} \lesssim 4 \times 10^{-3}.$

The approximate $\mu-\tau$ symmetry ($W_{\nu i}\cong W_{\tau i}$) of $[W_{\nu}(0)]$
is accounted for by the built-in $\mu-\tau$ symmetry of $[\widetilde{W}_{0}]$, $[\mathcal{B}]$, and $[\mathcal{C}]$.
Also, ($d,e$) are $\mu-\tau$ symmetry-breaking parameters.
As we shall see in the next section, as $A$ varies, $(d,e)$ remain small, while ($b,c$) will undergo
substantial changes.

Putting it all together, we have
\begin{equation}\label{para}
[W_{\nu}] = \left(\begin{array}{ccc}
  \smallskip
  \smallskip
   \frac{2}{3}+b-\frac{c}{2}, & \frac{1}{3}-b-\frac{c}{2}, & c \\ 
   \smallskip
   \smallskip
   \frac{1}{6}-\frac{b}{2}+\frac{c}{4}-\frac{d}{2}+e, & \frac{1}{3}+\frac{b}{2}+\frac{c}{4}-\frac{d}{2}-e, &\frac{1}{2}+d-\frac{c}{2} \\
   \smallskip
   \smallskip
   \frac{1}{6}-\frac{b}{2}+\frac{c}{4}+\frac{d}{2}-e, & \frac{1}{3}+\frac{b}{2}+\frac{c}{4}+\frac{d}{2}+e, & \frac{1}{2}-d-\frac{c}{2} \\
    \end{array}
    \right).
\end{equation}
The seeming complexity of $[W_{\nu}]$ is somewhat mitigated by its linear dependence
on the parameters, and manipulations on $[W_{\nu}]$ can be carried out without too much difficulty.
Another important feature is the strict correlation amongst the elements $W_{\alpha i}$.
For instance, the atmospheric neutrino measurement can determine $d^{2}$, leaving its sign ambiguous.
But once we choose a sign for $d$, it has to be used for all the elements $W_{\mu i}$ and $W_{\tau i}$.
Similarly, any error on one parameter would propagate to all $W_{\alpha i}$ elements with a
definite magnitude and sign.

The cofactor matrix, $[w_{\nu}]$, can be directly computed.  We find 
\begin{eqnarray}\label{cof}
[w_{\nu}]&=&\left(\begin{array}{ccc}
\smallskip
\smallskip
-\frac{7}{6}d-e, & \frac{5}{6}d-e, & -\frac{1}{6}d+e \\
 \smallskip
 \smallskip
 -\frac{1}{6}+\frac{3}{4}c+\frac{b}{2}+\frac{d}{3}, & \frac{1}{3}+\frac{b}{2}-\frac{3}{4}c-\frac{2}{3}d, &-\frac{1}{6}-b-\frac{d}{6}-e \\
\smallskip
\smallskip
  \frac{1}{6}-\frac{3}{4}c-\frac{b}{2}+\frac{d}{3}, & -\frac{1}{3}-\frac{b}{2}+\frac{3}{4}c-\frac{2}{3}d, &\frac{1}{6}+b-\frac{d}{6}-e \\
    \end{array}
    \right)   \nonumber \\
   & + &(c e-b d)
    \left(\begin{array}{ccc}
1& 1 & 1 \\
 1 & 1 &1 \\
   1 & 1 &1 \\
    \end{array}
    \right).
\end{eqnarray}

Using the matrices $[W_{\nu}]$ and $[w_{\nu}]$, we can readily express the variables ($x_{i},y_{j}$) in terms of the
set ($b,c,d,e$), which we will not write down explicitly here.

\section{Differential equations for matter effects}

When neutrinos propagate in a medium of constant density, their interactions induce a term in the
effective Hamiltonian, $H=\frac{1}{2E}M_{\nu}M^{\dag}_{\nu}$, given by 
\cite{Wolfenstein:1977ue, Mikheev:1986gs} 
$(\delta H)_{ee}=A=2\sqrt{2}G_{F}n_{e}E$. Thus, the neutrino mass eigenvalues 
squared ($D_{i}=m^{2}_{i}$) and mixing matrix are functions of $A$. 
It was shown \cite{Chiu:2010da}  
that they satisfy a set of differential equations, given by
\begin{equation}\label{dDA}
\frac{dD_{i}}{dA}=|V_{ei}|^{2}=W_{ei}, 
\end{equation}
\begin{equation}\label{dv}
\frac{dV_{\alpha i}}{dA} =\sum_{k\neq i} \frac{V_{\alpha k}V_{ei}}{D_{i}-D_{k}} V_{ek}^{*}.
\end{equation}
Here, Eq.~(\ref{dv}) is not rephasing invariant and should be used by making rephasing invariant
combinations constructed from $V_{\alpha i}$.  
This was done for the $(x_{i},y_{j})$ variables in Ref. \cite{Chiu:2012uc}, 
as well as for $W_{ei}$ and $w_{ei}$.
In this paper, we study the corresponding equations for $W_{\alpha i}=|V_{\alpha i}|^{2}$.  We find, from
Eq.~(\ref{dv}),
\begin{eqnarray}\label{dW}
\frac{d}{dA}W_{\alpha i} &=& \frac{d}{dA} (V^{*}_{\alpha i}V_{\alpha i}) \nonumber \\
&=& \sum_{k\neq i} \frac{1}{D_{i}-D_{k}}[V^{*}_{\alpha i}V_{\alpha k}V_{ei}V^{*}_{ek}+V_{\alpha i}V^{*}_{\alpha k}V^{*}_{ei}V_{ek}] \nonumber \\
&=&2\sum_{k\neq i}\frac{1}{D_{i}-D_{k}}Re(\Pi^{\alpha e}_{ik}),
\end{eqnarray}
where we have used the definitions
$\Pi^{\alpha e}_{ik}=V_{\alpha i}V^{*}_{\alpha k}V_{ek}V^{*}_{e i}$ and the relation
$(\Pi^{\alpha e}_{ik})=(\Pi^{\alpha e}_{ki})^{*}$.  We may further simplify the results by using
\begin{equation}\label{lamk}
\Lambda_{\gamma k}=Re(\Pi^{\alpha \beta}_{ij}), (\alpha\neq\beta \neq \gamma; i\neq j \neq k)
\end{equation}
Also, from the identity (which follows from $\sum_{\alpha}V^{*}_{\alpha j}V_{\alpha k}=\delta_{jk}$)
\begin{equation}
\sum_{\alpha}\Pi^{\alpha \beta}_{jk}=\delta_{jk}|V_{\beta k}|^{2},
\end{equation}
the relation
\begin{equation}
\Lambda_{\alpha i}+\Lambda_{\beta i}=-W_{\gamma j}W_{\gamma k}.
\end{equation}

We may now collect these results in a very compact form,
\begin{eqnarray}\label{dWdA}
\frac{1}{2} \frac{d}{dA}
\left(\begin{array}{ccc}
   W_{e1}& W_{e2} & W_{e3} \\
   W_{\mu1} & W_{\mu2} & W_{\mu3} \\
   W_{\tau1} & W_{\tau2} & W_{\tau3} \\
    \end{array}
    \right)
&=&\frac{1}{D_{1}-D_{2}}
\left(\begin{array}{ccc}
   W_{e1}W_{e2}, & -W_{e1}W_{e2},& 0\\
  \Lambda_{\tau3}, & -\Lambda_{\tau3}, & 0 \\
   \Lambda_{\mu3}, & -\Lambda_{\mu3}, & 0 \\
    \end{array}
    \right)  \nonumber \\ 
    &+&  \frac{1}{D_{2}-D_{3}}
    \left(\begin{array}{ccc}
   0, & W_{e2}W_{e3}, & -W_{e2}W_{e3} \\
   0, & \Lambda_{\tau1}, &-\Lambda_{\tau1} \\
    0, & \Lambda_{\mu1}, &-\Lambda_{\mu1} \\
    \end{array}
    \right)  \nonumber \\
& + & 
\frac{1}{D_{3}-D_{1}}
\left(\begin{array}{ccc}
   -W_{e1}W_{e3}, & 0, & W_{e1}W_{e3} \\
   -\Lambda_{\tau2}, & 0, & \Lambda_{\tau2} \\
    -\Lambda_{\mu2}, & 0, & \Lambda_{\mu2} \\
    \end{array}
    \right).
\end{eqnarray}   
The equation for $J$ was also computed \cite{Chiu:2011rj}. It reads 
\begin{equation}\label{dlnJ}
\frac{d}{dA}(\ln J)=\frac{-W_{e1}+W_{e2}}{D_{1}-D_{2}}+\frac{-W_{e2}+W_{e3}}{D_{2}-D_{3}}+\frac{-W_{e3}+W_{e1}}{D_{3}-D_{1}}.
\end{equation}


Eq.~(\ref{dWdA}) has a simple structure---it consists of pole terms with numerators being quadratic functions
of $W_{\alpha i}$.  In addition, with $e$ singled out by $(\delta H)_{ee} \neq 0$, it exhibits permutation
symmetry under the exchanges $\mu \leftrightarrow \tau$ and $i \leftrightarrow j \leftrightarrow k$.
This can be made explicit by rewriting Eq.~(\ref{dWdA}) in the form,
\begin{eqnarray}
\frac{1}{2}\frac{d}{dA}W_{ei}&=&\sum_{k\neq i} \frac{W_{ei}W_{ek}}{D_{i}-D_{k}} \nonumber \\
\frac{1}{2}\frac{d}{dA}W_{\alpha i}&=&\sum_{k\neq i \neq j} \frac{\Lambda_{\beta j}}{D_{i}-D_{k}} , \alpha \neq \beta, (\alpha,\beta)=(\mu,\tau),
\end{eqnarray}
which are manifestly invariant in form under the exchanges $(\alpha \leftrightarrow \beta)$ and $(i \leftrightarrow j \leftrightarrow k)$.
(Note that $\Lambda_{\beta j}$ transforms like $W_{\beta j}$ under permutation, according to Eq.~(\ref{lambda}).)
This symmetry greatly constrains the form of the evolution equations. 
(See also Eqs.~(\ref{prob}) and (\ref{aa}) in Sec. VI, where the probability functions are clearly invariant in form
under permutations of the indices.)
In contrast to the SP, where $\theta_{ij}$,
despite their appearances, have complicated permutation properties. 
For example, under the exchange $2 \leftrightarrow 3$, the corresponding transformation is not $\theta_{12} \leftrightarrow \theta_{13}$,
owing to the noncommutativity of the submatrices which combine into the mixing matrix.
One would thus not expect simple structures for $(d/dA)\theta_{ij}$.

From these equations, we can read off a number of interesting properties.  Thus, from Eqs.~(\ref{dDA}) and (\ref{dlnJ}),
we infer the ``matter invariant" \cite{Chiu:2010da,Toshev:1991ku,Naumov:1993vz,Harrison:1999df},
\begin{equation}
\frac{d}{dA}\ln[J(D_{1}-D_{2})(D_{2}-D_{3})(D_{3}-D_{1})]=0.
\end{equation}
Another ``matter invariant" follows immediately from Eqs.~(\ref{dWdA}) and (\ref{dlnJ}),
\begin{equation}
\frac{d}{dA}[J^{2}/(W_{e1}W_{e2}W_{e3})]=0.
\end{equation}

We turn now to a more detailed analysis of these differential equations. We note first that the group
$(W_{ei},D_{i}), (i=1,2,3)$, according to Eqs.~(\ref{dDA}) and (\ref{dWdA}), forms a closed set (see also 
Eq.~(\ref{dlnJ}) for $J$.  With the known vacuum values of $W_{ei}$ and $(D_{i}-D_{j})$,
we can thus solve for these parameters as functions of $A$.  Armed with these results,
the remaining elements $W_{\alpha i}$ ($\alpha=\mu,\tau$, and $i=1,2,3$) can be analyzed. 

Consider explicitly the equations for ($W_{e1},W_{e2},D_{2}-D_{1}$),
\begin{equation}\label{dD21}
\frac{d}{dA}(D_{2}-D_{1})=W_{e2}-W_{e1},
\end{equation}
\begin{equation}\label{dWe1}
\frac{1}{2}\frac{d}{dA}W_{e1}=-\frac{W_{e1}W_{e2}}{D_{2}-D_{1}}-\frac{W_{e1}W_{e3}}{D_{3}-D_{1}},
\end{equation}
\begin{equation}\label{dWe2}
\frac{1}{2}\frac{d}{dA}W_{e2}=\frac{W_{e1}W_{e2}}{D_{2}-D_{1}}-\frac{W_{e2}W_{e3}}{D_{3}-D_{2}}
\end{equation}
For small $A$ values, the term $\propto 1/(D_{2}-D_{1})$ dominates, since here $D_{3} \gg (D_{2},D_{1})$.
In addition, $W_{e3} \ll 1$, so that there is a ``double suppression" for the second term in Eqs.~(\ref{dWe1},\ref{dWe2}),
which can be well-approximated by
\begin{equation}\label{dWe1App}
-\frac{dW_{e1}}{dA}=\frac{dW_{e2}}{dA}=\frac{2W_{e1}W_{e2}}{D_{2}-D_{1}}.
\end{equation}
Eqs.~(\ref{dD21}) and (\ref{dWe1App}) are exactly those for a level-crossing problem for two flavors.
The solutions, as given in Eq. (22) of Ref. \cite{Chiu:2011rj}, 
with the approximate initial conditions $W_{e1}(0)=2/3$, $W_{e2}(0)=1/3$, 
$\delta m^{2}_{21}=\delta_{0}=7.53 \times 10^{-5} eV^{2}$, plus the definition $\Delta_{21}=D_{2}-D_{1}$, are
\begin{equation}\label{eq21}
\Delta_{21}=[A^{2}-\frac{2}{3}\delta_{0}A+\delta_{0}^{2}]^{1/2},
\end{equation}
\begin{equation}\label{eq22}
W_{e1}=\frac{1}{2}[1-(A-\frac{1}{3}\delta_{0})/\Delta_{21}],
\end{equation}
\begin{equation}\label{eq23}
W_{e2}=\frac{1}{2}[1+(A-\frac{1}{3}\delta_{0})/\Delta_{21}].
\end{equation}
Note also that $(d/dA)(W_{e1}W_{e2}\Delta^{2}_{21})=0$.  Thus, we find a typical resonance behavior
near the (lower) resonance point, $A=A_{l}\cong \delta_{0}/3$.  Away from $A_{l}$,
$\Delta_{21}\simeq D_{2} \rightarrow A$, $W_{e2} \rightarrow 1$, and 
$W_{e1} \rightarrow W^{0}_{e1}W^{0}_{e2}/\Delta^{2}_{21} \sim 2/(9A^{2})$.
Notice the effects of decoupling.  As $A$ pulls away from $A_{l}$, the state $|2\rangle$ approaches
a pure $|e\rangle$ state.  All the parameters tend to their limiting values of no mixing, 
independent of their initial configurations.

Similarly (for normal ordering), as $A$ increases further, when $D_{32}$ reaches a minimum near
$D_{2}\simeq A \sim m^{2}_{3}$, we have the equations,
\begin{equation}
\frac{d}{dA}(D_{3}-D_{2})=W_{e3}-W_{e2},
\end{equation}
\begin{equation}\label{dWe2App}
\frac{1}{2}\frac{d}{dA}W_{e2}=-\frac{W_{e2}W_{e3}}{D_{3}-D_{2}}+\frac{W_{e1}W_{e2}}{D_{2}-D_{1}}
\cong -\frac{W_{e2}W_{e3}}{D_{3}-D_{2}},
\end{equation}
\begin{equation}\label{dWe3App}
\frac{1}{2}\frac{d}{dA}W_{e3}=\frac{W_{e2}W_{e3}}{D_{3}-D_{2}}-\frac{W_{e1}W_{e3}}{D_{3}-D_{1}}
\cong \frac{W_{e2}W_{e3}}{D_{3}-D_{2}}.
\end{equation}
The initial conditions for these equations are the values of $W_{ei}$ and $D_{i}$ obtained for $A\gg A_{l}$.
Again, the dropped terms are doubly suppressed from $(D_{2},D_{3}) \gg D_{1}$ and $W_{e1} \rightarrow 0$.
So now we have another (higher) resonance near $A\simeq \Delta_{0}=2.45 \times 10^{-3} eV^{2} \cong 31 \delta_{0}$,
\begin{equation}\label{eq27}
\Delta_{32}=[A^{2}-2\Delta_{0}A+\Delta_{0}^{2}]^{1/2},
\end{equation}
\begin{equation}
W_{e2}=\frac{1}{2}[1-(A-q_{h}\Delta_{0})/\Delta_{32}],
\end{equation}
\begin{equation}\label{eq29}
W_{e3}=\frac{1}{2}[1+(A+q_{h}\Delta_{0})/\Delta_{32}],
\end{equation}
where $q_{h}\cong 1-2W_{e3}(0)$.  Note that the contributions from the pole term $\propto 1/(D_{3}-D_{1})$,
according to Eq.~(\ref{dWdA}), are always doubly suppressed.
First, from the denominator, and second, from $W_{e1}W_{e3}\ll 1$, for all $A$ values.

In summary, the set $(W_{ei},\Delta_{ij})$, according to Eqs.~(\ref{eq21}-\ref{eq23}) and (\ref{eq27}-\ref{eq29}), 
can be described in terms of two well separated level-crossing problems.
While the mass eigenvalues $D_{i}$ take turns to rise proportionally to $A$,
the $W_{ei}$ change rapidly only near $A_{l}$ ($\cong \delta_{0}/3$, the ``lower resonance") and
$A_{h} (\cong \Delta_{0}\cong 31 \delta_{0}$, the ``higher resonance").
There are two regions 1) $A_{i}$, with $A_{l}<A_{i}<A_{h}$; and 2) $A_{d}$, with $A_{d} \gg A_{h}$, in which
$W_{ei}$ are stationary.  The span of $A_{i}$ and $A_{d}$ can be obtained from the 
positions and widths of the two resonances.
A conservative estimate yields: $5\delta_{0} \lesssim A_{i} \lesssim 15 \delta_{0}$, $A_{d} \gtrsim 50 \delta_{0}$.

In terms of the parameters $b$ and $c$, we see that $b\rightarrow -\frac{2}{3}+\frac{c_{0}}{2}$ and $c=c_{0}$ 
as $A\rightarrow A_{i}$, so that $W_{e1} \rightarrow 0$ and $W_{e2} \rightarrow 1-c_{0}$.
After the higher resonance, $A \rightarrow A_{d}$, $c\rightarrow 1$, $b\rightarrow -1/2$.

We now turn to the matrix elements $W_{\mu i}$ and $W_{\tau i}$.  Up to the region $A \sim A_{i}$,
from Eq.~(\ref{dWdA}), for $W_{\mu 3}$ and $W_{\tau 3}$, 
there is no contribution from the dominant pole term ($\propto 1/(D_{2}-D_{1})$),
contributions from the other pole terms ($\propto 1/(D_{3}-D_{1})$ and $1/(D_{3}-D_{2})$) are also
doubly suppressed (see Eq.~(\ref{LambdaAi}) below).  Thus, to a very good approximation,  
\begin{equation}
d(A_{i})=d_{0}.
\end{equation}  
From $W_{\mu 2}=\frac{1}{2}W_{e1}+\frac{c}{2}-\frac{d}{2}-e$, with $W_{e1} \rightarrow 0$, we find
\begin{equation}
W_{\mu 2} (A_{i})=\frac{c_{0}}{2}-\varepsilon_{0}, \hspace{0.2in} \varepsilon_{0}= \frac{d}{2}+e.
\end{equation}
Similarly,
\begin{equation}
W_{\tau 2}(A_{i})=\frac{c_{0}}{2}+\varepsilon_{0}.
\end{equation}
And, positivity demands
\begin{equation}
|\varepsilon_{0}|<\frac{c_{0}}{2},
\end{equation}
i.e., $e(A_{i}) \rightarrow -(d_{0}/2)+\varepsilon_{0}$.  This also fixes the elements $W_{\mu 1}$ and $W_{\mu 2}$.
To within an accuracy of $0.01$, we may ignore $\varepsilon_{0}$ and obtain
\begin{equation}\label{WAi}
[W(A_{i})] \cong \left(\begin{array}{ccc}
   0, & 1-c_{0}, & c_{0} \\
   \smallskip
   \frac{1}{2}-d_{0}, & c_{0}/2, & \frac{1}{2}-(c_{0}/2)+d_{0} \\
   \smallskip
    \frac{1}{2}+d_{0}, & c_{0}/2, & \frac{1}{2}-(c_{0}/2)-d_{0} \\
    \end{array}
    \right).
  \end{equation}
This shows that, after the lower resonance, the matrix $W$ assumes a very simple form,
depending only on the small vacuum parameters ($c_{0},d_{0}$).  As a check
on the stability of $W$ over the range $A_{i}$, we calculate from $W(A_{i})$
\begin{equation}\label{LambdaAi}
2[\Lambda(A_{i})] = \left(\begin{array}{ccc}
   c_{0}(1-c_{0})/2, & -\frac{1}{2} (1-c_{0})+2d^{2}_{0}, & -c_{0}/2 \\
   \smallskip
  -c_{0}(1-c_{0})+c_{0}d_{0}, & c_{0}d_{0}, & -c_{0}d_{0} \\
   \smallskip
     -c_{0}(1-c_{0})-c_{0}d_{0}, & -c_{0}d_{0}, & c_{0}d_{0} \\
    \end{array}
    \right).
  \end{equation}
That all elements $\Lambda_{\mu i}$ and $\Lambda_{\tau i}$ are small is consistent with the constancy
of $W_{\mu i}$ and $W_{\tau i}$.

  \begin{figure}[ttt]
\caption{The evolution of $D_{1}$ (dotted), $D_{2}$ (solid), 
and $D_{3}$ (dot-dashed) for the $\nu$ sector in matter under normal (left and middle) and inverted (right) orderings.} 
\centerline{\epsfig{file=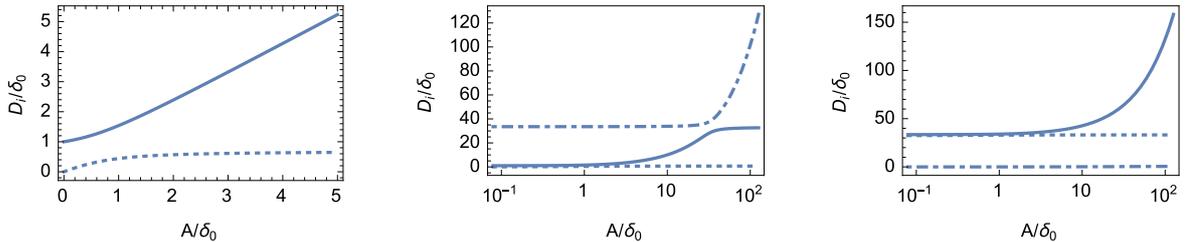,width= 16cm}}
\end{figure} 

Finally, after the higher resonance (for NO), $A \sim A_{d}$, $W_{\alpha 1}$ is unchanged, while
$W_{e3} \rightarrow 1$ and $W_{\mu 3}$ and $W_{\tau 3} \rightarrow 0$, but
($W_{\mu 2}+W_{\mu 3}$) and ($W_{\tau 2}+W_{\tau 3}$) are stationary.
This means that $b(A_{d}) \rightarrow -1/6$, $c(A_{d}) \rightarrow 1$, $d(A_{d}) \rightarrow 0$,
$e(A_{d}) \rightarrow -d_{0}$, and  
\begin{equation}\label{WAd}
[W(A_{d})] = \left(\begin{array}{ccc}
   0, & 0, &1 \\
   \smallskip
  \frac{1}{2}-d_{0}, & \frac{1}{2}+d_{0}, &0 \\
   \smallskip
     \frac{1}{2}+d_{0}, & \frac{1}{2}-d_{0}, &0 \\
    \end{array}
    \right).
  \end{equation}

To summarize, Eq.~(\ref{dDA}) and Eq.~(\ref{dWdA}), with the known (approximate) vacuum values
($\delta_{0},\Delta_{0},W^{0}_{e1},W^{0}_{e3}$) as initial conditions, turn out to have very good 
approximate solutions for all $A$ values. The mass eigenvalues squared, $D_{i}$, rise proportionally to $A$, successively.
The mixing matrix, $[W(A)]$, has two well-separated regions (around $A_{l}$ and $A_{h}$) in which
some matrix elements evolve rapidly.  There are two regions, $A \sim A_{i}$ and $A \gtrsim A_{d}$,
wherein all the matrix elements are nearly stationary. These matrices are given in Eqs.~(\ref{WAi}) and (\ref{WAd}).
It is interesting to note that, while there are four parameters ($b_{0},c_{0},d_{0},e_{0}$) in the 
vacuum mixing matrix ($[W(0)]$), for $A \sim A_{i}$, $[W(A_{i})]$ depends only on two, ($c_{0},d_{0}$).
Finally, for large $A$ values, the only parameter in $[W(A_{d})]$ is $d_{0}$.
It is also interesting to note that, from the ``matter invariant", $[J^{2}/(W_{e1}W_{e2}W_{e3}]$,
and the rough estimate, $W_{e1}W_{e2}W_{e3} \sim 1/A^{2}$, the CP-violating effects
are suppressed as $A$ increase, both for the normal ordering (NO) and the inverted ordering (IO). 

In addition to $[W(0)]$ (Eq.~(\ref{para})), $[W(A_{i})]$ (Eq.~(\ref{WAi})), and $[W(A_{d})]$ (Eq.~(\ref{WAd}))
under NO, we show in Appendix B the possible parametrization of them under IO. Furthermore, the resultant $[\Lambda]$ matrices are also presented.


\section{Numerical solutions}

 \begin{figure}[ttt]
\caption{Numerical solutions of $W_{\alpha i}$ for the $\nu$ sector as functions of $A/\delta_{0}$ under the normal ordering (solid) and
the inverted ordering (dashed). } 
\centerline{\epsfig{file=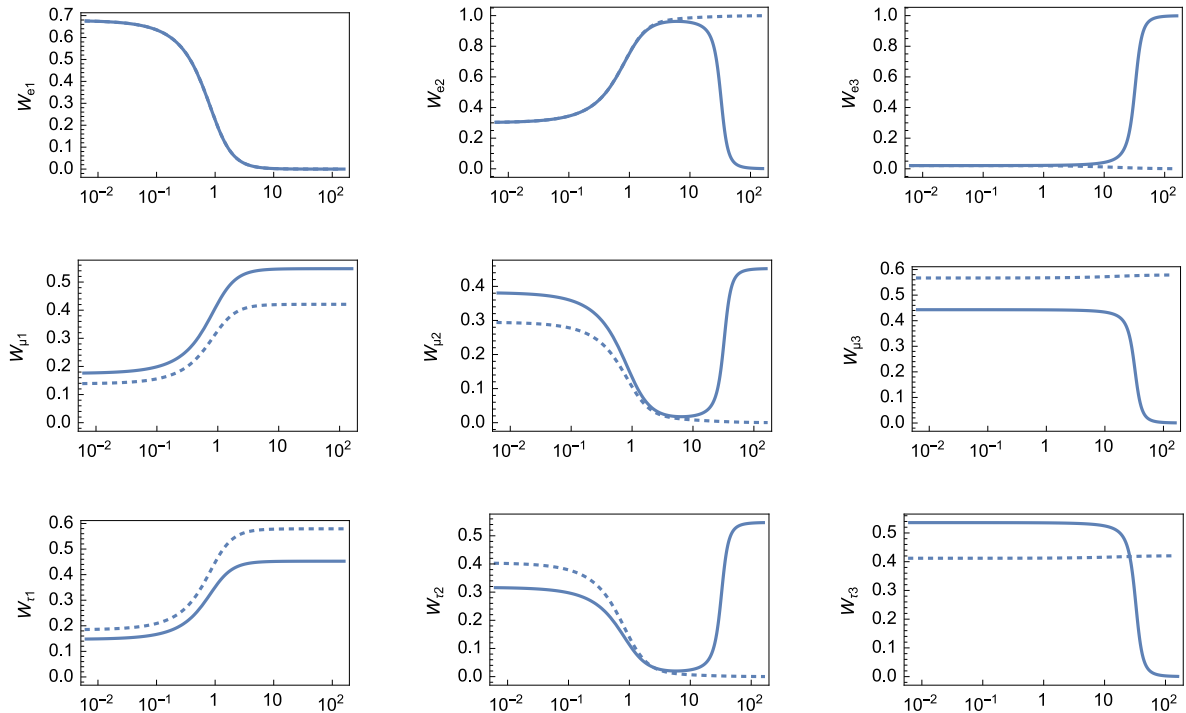,width= 16cm}}
\end{figure} 

The general features of $D_{i}$ and $W_{\alpha i}$ for the $\nu$ sector
in matter are plotted as functions of $A/\delta_{0}$ in Fig. 1 and Fig. 2, respectively, 
under both the normal (NO) and the inverted (IO) orderings. 
It is seen that $D_{i}$ goes through both lower and higher resonances under NO,
while there is no resonance under IO.
The elements $W_{\alpha i}$ may go through the resonance at $A \lesssim \delta_{0}$ or
$A \sim \Delta_{0}$, or both, depending on the neutrino types ($\nu$ or $\bar{\nu}$)
and the mass orderings (NO or IO).
We will not show the plots for the $\bar{\nu}$ sector, in which there is a higher resonance
for $\bar{D_{i}}$ under IO.  The behavior of $\overline{W}_{\alpha i}$ for the $\bar{\nu}$ sector
can be summarized as follows: 
(i) For NO, there is no resonance. 
(ii) For IO, only $\overline{W}_{\alpha 1}$ and $\overline{W}_{\alpha 3}$
go though the higher resonance. 
It should be emphasized that at the present accuracy, it is unlikely to reach the vacuum values of all the $[W_{\nu}]$ elements to the same satisfactory level.  Thus, for illustration purpose only, we roughly estimate the values of
$W_{\alpha i}$ in vacuum based on the available analyses
(see, e.g., \cite{Forero:2014bxa,Gonzalez-Garcia:2015qrr,Capozzi:2016rtj}), 
\begin{equation}\label{Winitial-NO}
[W_{\nu}]_{N} \approx \left(\begin{array}{ccc}
   0.678, & 0.301, & 0.021 \\
   0.175, & 0.382, & 0.443 \\
   0.147, & 0.317, & 0.536 \\
    \end{array}
    \right)
 \end{equation}
under NO, and 
\begin{equation}\label{Winitial-IO}
[W_{\nu}]_{I} \approx \left(\begin{array}{ccc}
   0.678, & 0.301, & 0.021 \\
   0.138, & 0.295, & 0.567 \\
   0.184, & 0.404, & 0.412 \\
    \end{array}
    \right)
  \end{equation}
under IO.


  \begin{figure}[ttt]
\caption{Numerical solutions of $\Lambda_{\gamma k}$ for the $\nu$ sector as functions of $A/\delta_{0}$ under the normal ordering (solid) and the inverted ordering (dashed).} 
\centerline{\epsfig{file=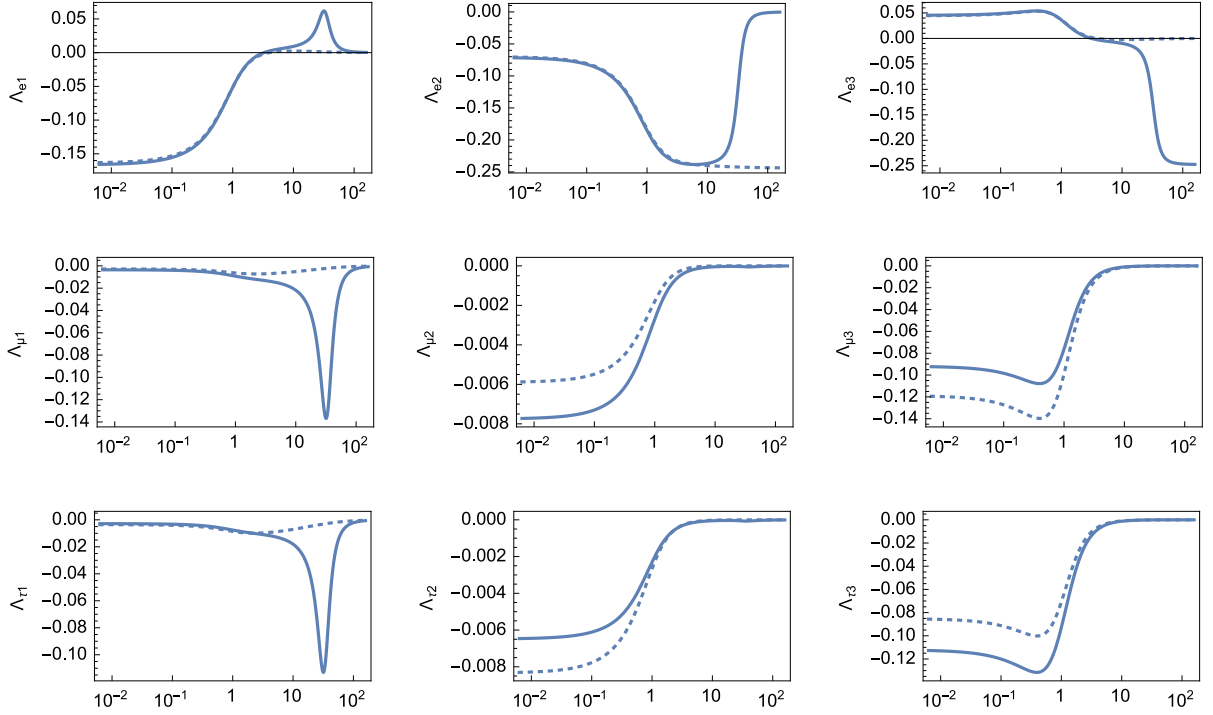,width= 16cm}}
\end{figure} 


As will be seen in the next section, the quantities $\Lambda_{\gamma k}$ defined in Eq.~(\ref{lamk}),
$\Lambda_{\gamma k}=Re[V_{\alpha i}V_{\beta j}V^{*}_{\alpha j}V^{*}_{\beta i}]=Re[\Pi^{\alpha \beta}_{ij}]$,
play important roles in the transition probability for neutrino oscillation. They represent the relative weight of each 
$\sin^{2}\Phi_{ij}$ component in the probability function. Explicitly,
\begin{eqnarray}\label{Lambda}
\Lambda_{e 1}&=&\frac{1}{2}(W_{\mu 2}W_{\tau 3}+W_{\mu 3}W_{\tau 2}-W_{e1}),    \nonumber  \\
\Lambda_{e 2}&=&\frac{1}{2}(W_{\mu 1}W_{\tau 3}+W_{\mu 3}W_{\tau 1}-W_{e2}),   \nonumber \\
\Lambda_{e 3}&=&\frac{1}{2}(W_{\mu 1}W_{\tau 2}+W_{\mu 2}W_{\tau 1}-W_{e3}),   \nonumber \\
\Lambda_{\mu 1}&=&\frac{1}{2}(W_{e3}W_{\tau 2}+W_{e2}W_{\tau 3}-W_{\mu 1}),    \nonumber  \\
\Lambda_{\mu 2}&=&\frac{1}{2}(W_{e1}W_{\tau 3}+W_{e3}W_{\tau 1}-W_{\mu 2}),   \nonumber \\
\Lambda_{\mu 3}&=&\frac{1}{2}(W_{e2}W_{\tau 1}+W_{e1}W_{\tau 2}-W_{\mu 3}),   \nonumber \\
\Lambda_{\tau 1}&=&\frac{1}{2}(W_{e3}W_{\mu 2}+W_{e2}W_{\mu 3}-W_{\tau 1}), \nonumber \\
\Lambda_{\tau 2}&=&\frac{1}{2}(W_{e3}W_{\mu 1}+W_{e1}W_{\mu 3}-W_{\tau 2}), \nonumber \\
\Lambda_{\tau 3}&=&\frac{1}{2}(W_{e1}W_{\mu 2}+W_{e2}W_{\mu 1}-W_{\tau 3}).
\end{eqnarray}
We plot $\Lambda_{\gamma k}$ as functions of $A/\delta_{0}$ in Fig. 3.

\section{Applications to the Experiments}

As the neutrinos travel through a baseline, the flavor transition probability
is given by the well-known expression,
\begin{eqnarray}\label{prob}
P(\nu_{\alpha} \rightarrow \nu_{\beta})&=&\delta _{\alpha \beta}-4 \sum_{j>i} Re(V_{\alpha i}V_{\beta j}V^{*}_{\alpha j}V^{*}_{\beta i})\sin^{2}(\Phi_{ji})  \nonumber \\
&+&2\sum_{j>i}Im(V_{\alpha i}V_{\beta j}V^{*}_{\alpha j}V^{*}_{\beta i})\sin(2\Phi_{ji}),
\end{eqnarray}
where the explicit form of $Re[V_{\alpha i}V_{\beta j}V^{*}_{\alpha j}V^{*}_{\beta i}]=Re[\Pi^{\alpha \beta}_{ij}]=\Lambda_{\gamma k}$
are given by Eq.~(\ref{Lambda}), 
$Im(V_{\alpha i}V_{\beta j}V_{\alpha j}^{*}V_{\beta i}^{*})=(\epsilon_{\alpha \beta \gamma})(\epsilon_{ijk})J$,
$\Phi_{ji}\equiv (D_{j}-D_{i})L/(4E)$, $L$ is the baseline length, and $E$ is the neutrino energy.
More explicitly, for the disappearance channel, 
\begin{equation}\label{aa}
P(\nu_{\alpha} \rightarrow \nu_{\alpha})=1-4(W_{\alpha 1}W_{\alpha 2}\sin^{2}\Phi_{21}+W_{\alpha 1}W_{\alpha 3}
\sin^{2}\Phi_{31}+W_{\alpha 2}W_{\alpha 3}\sin^{2}\Phi_{32}),
\end{equation}
and for the appearance channel, $\alpha \neq \beta \neq \gamma$,
\begin{eqnarray}\label{ab}
P(\nu_{\alpha}\rightarrow \nu_{\beta})=&-&4[\Lambda_{\gamma 3}\sin^{2}\Phi_{21} 
+ \Lambda_{\gamma 2}\sin^{2}\Phi_{31} 
+ \Lambda_{\gamma 1}\sin^{2}\Phi_{32}] \nonumber \\
&+& 2J[\sin 2\Phi_{21}-\sin 2\Phi_{31} + \sin 2\Phi_{32}].
\end{eqnarray}

For neutrinos in vacuum, the $[W_{\nu}]$ matrix is parametrized by Eq.~(\ref{para}), and the values of $\Lambda_{0}=\Lambda (0)$
are given by
\begin{equation}\label{LambdaA0}
2[\Lambda(0)] \cong \left(\begin{array}{ccc}
  \smallskip
   -\frac{1}{3}-\frac{1}{2}b_{0}+\frac{5}{12}c_{0}, &-\frac{1}{6}+\frac{1}{2}b_{0}+\frac{7}{12}c_{0}, 
   &\frac{1}{9}-\frac{3}{4}c_{0}-\frac{1}{6}b_{0}  \\
  
   \smallskip
  
  -\frac{1}{3}c_{0}+\frac{1}{6}d_{0}-e_{0}, & -\frac{2}{3}c_{0}-\frac{1}{6}d_{0}+e_{0}, 
  & -\frac{2}{9}+\frac{1}{3}b_{0}+\frac{1}{2}c_{0}-\frac{1}{2}d_{0}+\frac{1}{3}e_{0} \\
   \smallskip
   
    -\frac{1}{3}c_{0}-\frac{1}{6}d_{0}+e_{0}, &-\frac{2}{3}c_{0}+\frac{1}{6}d_{0}-e_{0}, 
   & -\frac{2}{9}+\frac{1}{3}b_{0}+\frac{1}{2}c_{0}+\frac{1}{2}d_{0}-\frac{1}{3}e_{0}  \\
    \end{array}
    \right)
  \end{equation}
where the quadratic terms are ignored.  Note that $\Lambda_{\mu i}$ and $\Lambda_{\tau i}$ are related by 
the replacement $d_{0} \rightarrow -d_{0}$ and $e_{0} \rightarrow -e_{0}$.

For neutrinos in matter, it is seen from Fig. 1 that the values of some of the $W_{\alpha i}$ can vary significantly.  
Hence, one would expect  to see a sizable impact 
to the analysis of long baseline experiments (LBL)
(for an incomplete list, see, e.g. \cite{LBL-1,Bandyopadhyay:2007kx}),  
many of which operate in the range corresponding to $A\approx 5\delta_{0}$ to $15 \delta_{0}$,
so that $A\sim A_{i}$.
With the explicit expression of $\Lambda (A_{i})$, e.g., under NO in Eq.~(\ref{LambdaAi}), 
the transition probabilities, Eqs.~(\ref{aa}) and (\ref{ab}), would contain $b_{0}$, $c_{0}$, and the undetermined $d_{0}$ for a detailed analysis of the LBL data.


Using Eq.~(\ref{LambdaAi}),
it can be verified that with $A\sim A_{i}$, the appearance channels, $\nu_{e} \rightarrow \nu_{\mu}$ and
$\nu_{e} \rightarrow \nu_{\tau}$, are insignificant and their probabilities are only of order $\sim c_{0}$ or less.
In  addition, for $P(\nu_{\mu}\rightarrow \nu_{\tau})$, the contribution from the 
dominant term ($\sin^{2}\Phi_{31}$) is of order $\sim 1$, while that from $\sin^{2}\Phi_{21}$ and 
$\sin^{2}\Phi_{32}$ terms are only of order $\sim c_{0}^{2}$. 
Thus, the appearance channel $\nu_{\mu}\rightarrow \nu_{\tau}$ could be significant at $A\sim A_{i}$ 
if the experimental setup is properly chosen so that $\sin^{2}\Phi_{31}$ is large.
On the other hand in $P(\nu_{\mu}\rightarrow \nu_{e})$
the contribution from $\sin^{2}\Phi_{31}$ and $\sin^{2}\Phi_{21}$ terms become of order $\sim c_{0}d_{0}$, 
while that from $\sin^{2}\Phi_{32}$ terms are of order $\sim c_{0}$.

In addition to the experiments involving terrestrial neutrino sources, intensive effort has also been
devoted to the study of extraterrestrial neutrino sources such as the neutrinos 
from a core-collapsed supernova \cite{Kuo:1989qe,Dighe:1999bi}.  
One of the characteristics of these neutrino fluxes is 
that they travel through a very dense media before they exit. 
Thus, the induced mass corresponds to $A\gtrsim A_{d}$.  Our results are therefore relevant
to such processes, especially in regard to the question of NO vs IO.


By using the $W$-centric parametrization, 
a proper estimation of $W_{\alpha i}$ in matter leads to simple expressions of $\nu$ oscillation
probabilities to within the accuracy of $\sim 0.01$.
The expressions reveal explicitly the relative order of magnitude of the contributions from $\sin^{2}\Phi_{ij}$.
Thus, by choosing a proper experimental setup which leads to a significant magnitude of $\sin^{2}\Phi_{ij}$, 
a careful analysis of data may shed 
some light on the parameter $d_{0}$, which is closely related to
the $\mu-\tau$ asymmetry, $J$, and $\theta_{23}$.
We shall leave the detailed analysis to a future work.


\section{conclusion}
 
The central issue in flavor physics is the determination of the mixing matrices of
quarks and neutrinos; or rather because of the rephasing invariance, the measurement of the absolute values 
of their matrix elements.  In the quark sector, this effort has culminated in extremely accurate results for the squared 
CKM matrix elements (to order $10^{-5}$), which will be summarized in Appendix A.
For the neutrino sector, despite its ``new comer" status, our knowledge on $|V_{\alpha i}|^{2}$ 
is nevertheless quite substantial. Of the elements of $[W_{\nu}]$ ($W_{\alpha i}=|V_{\alpha i}|^{2}$),
five are rather accurately known.  In this paper we introduce a parametrization of $[W_{\nu}]$ in which
the unitarity conditions are strictly imposed.  
This brings out explicitly the strong correlations between the elements
of $[W_{\nu}]$.  A precision measurement on one of the lessor known elements would go a long way toward
fixing the whole matrix.

Another interesting subject is the study of neutrino propagation in matter, 
for which its parameters become
functions of $A$, the induced neutrino mass.  In this paper, we derive a set of differential equations 
obeyed by the elements $W_{\alpha i}$.  The distinctive feature of these equations is their 
dependence on the variables $\Lambda_{\gamma k}$, which also play central roles in the 
formulas of neutrino oscillation probabilities, in addition to the renormalization group 
equations of quarks and neutrinos. Note also that $\Lambda_{\gamma k}$ are simple functions of
rephasing invariant variables ($W_{\alpha i}$ or ($x_{i},y_{j}$)), instead of their complicated forms in the SP.  
Thus, it would be worthwhile to reanalyze the experiments directly using 
rephasing invariant variables so as to avoid
losing information in translation.

As for solving these differential equations, it is found, somewhat fortuitously, that they have very good 
approximate solutions for all values of $A$, when the initial conditions are taken to be the currently available
(albeit incomplete) values for vacuum neutrino parameters.  
The results (for NO) are dominated by two well-separated
level-crossing (resonance) solutions.  Outside of these resonance regions, 
all the mixing parameters are  nearly stationary.
It is noteworthy that several LBL experiments operate in the $A$ range which coincides with
the stability region.  This situation should be helpful in deciphering the implications 
of the experimental results.

\acknowledgments                 
SHC was supported by the Ministry of Science and Technology of Taiwan, 
Grant No. MOST 104-2112-M-182-004.

\appendix

\section{Estimation of $[W_{\alpha i}]$ for quark mixing}

Although the $W$-centric parametrization has its advantage over the standard parametrization in some aspects,
current measurements of quark and neutrino mixings are all based on the standard parametrization.
Thus, it would be useful to derive values of the $[W]$ matrix elements for both
the quark and the neutrino mixings.

With the currently available precision measurement of the elements of $V_{CKM}$, it would be useful to derive
$[W]_{\alpha i}$, $[w]_{\alpha i}$, $(x_{i},y_{j})$, and their respective uncertainties for the quark mixing.
Based on the $|V_{\alpha i}|$ values given by the Particle Data Group \cite{PDG}, 
we obtain $[W_{\alpha i}]=|V_{\alpha i}|^{2}$,
\begin{equation}\label{appW}
[W]=\left(\begin{array}{ccc}
   0.94934\pm0.00023, & 0.05065 \pm 0.00023, & (1.3 \pm 0.1)\times 10^{-5} \\
   0.05059 \pm 0.00023, & 0.94772 \pm 0.00025, & (1.69 \pm 0.11)\times 10^{-3} \\
   (7.7 \pm 0.6)\times 10^{-5}, & (1.62 \pm 0.11)\times10^{-3}, & 0.99830 \pm 0.00010 \\
    \end{array}
    \right).
  \end{equation}
For $w_{\alpha i}$, one obtains
\begin{equation}\label{appww}
[w]=\left(\begin{array}{ccc}
   0.94611\pm0.00027, & -0.05050 \pm 0.00022, & (0.95 \pm 0.75)\times 10^{-5} \\
   -0.05056 \pm 0.00023, & 0.94773 \pm 0.00025, & (-1.53 \pm 0.10)\times 10^{-3} \\
   (7.4 \pm 0.6)\times 10^{-5} ,& (-1.60 \pm 0.10)\times10^{-3}, & 0.89715 \pm 0.00032 \\
    \end{array}
    \right).
  \end{equation}

To derive the values of $(x_{i},y_{j})$, we note that each parameter can be calculated
in three ways, using Eqs.~(\ref{appW}) and ~(\ref{appww}). Since the three values have very
different standard deviations, we take a weighted average (see, e.g., PDG 2016 booklet \cite{PDG}, Eq. (39.8)).
For example, for $x_{2}$, which is dominated by $(1/2)(W_{td}+w_{td})$, we find $x_{2}=(7.5 \pm 0.4)\times 10^{-5}$.
For $y_{3}$, the same estimate yields $y_{3}=(-0.16 \pm 0.27) \times 10^{-5}$.
Note that with the known values of $(x_{1},x_{2},x_{3},y_{1},y_{2})$,   
$y_{3}$ can also be calculated from $J^{2}=x_{1}x_{2}x_{3}-y_{1}y_{2}y_{3}$, and the measured value
$J=(3.04 \pm 0.21) \times 10^{-5}$.  This yields $y_{3}=(-0.17 \pm 0.37) \times 10^{-5}$.  
We can now list our best estimation of $(x_{i},y_{j})$,
\begin{eqnarray}
x_{1}=0.94772 \pm 0.00010,   \hspace{0.2in} y_{1}=(-1.61 \pm 0.05)\times 10^{-3} \nonumber \\
x_{2}=(7.5 \pm 0.4) \times 10^{-5},    \hspace{0.2in} y_{2}=-0.05058 \pm 0.00010 \nonumber \\
x_{3}= (1.1\pm 0.4) \times 10^{-5} \hspace{0.2in}  y_{3}= (-0.16 \pm 0.32)\times 10^{-5}.  
\end{eqnarray}
With the precision measurements of $|V_{CKM}|$, it is seen that $(x_{i},y_{j})$ can be determined consistently,
although some of them have rather large uncertainties.  It would be interesting to analyze the experimental data
directly in terms of $(x_{i},y_{j})$.  One would expect to have more accurate results without having
to rely on the intermediaries like $W_{\alpha i}$ and $w_{\alpha i}$.

It is tempting to follow the same methodology in order to convert the known $[W_{\nu}]$ elements
into values of $(x_{i},y_{j})$.  However, at the present level of accuracy, a consistent solution
is not available.

While the $(x_{i},y_{j})$ parametrization is applicable in general, for quark mixing, we may also
use another parametrization that incorporates the feature of $|V_{CKM}|$, similar to Eq.~(\ref{nuW})
for $[W_{\nu}]$.  We write
\begin{equation}
[W_{q}]=[W^{0}_{CKM}]+p[\mathcal{P}]+q[\mathcal{Q}]+r[\mathcal{R}]+s[\mathcal{S}].
\end{equation}
Here
\begin{equation}\label{quarkW}
[W^{0}_{CKM}]=\left(\begin{array}{ccc}
 1,& 0, & 0 \\
 0, & 1, &0 \\
    0, & 0, &1 \\
    \end{array}
    \right) ,
\end{equation}
and 
\begin{eqnarray}
[\mathcal{P}]=\left(\begin{array}{ccc}
   1, & -1, & 0 \\
-1, & 1, & 0 \\
    0, & 0,& 0 \\
    \end{array}
    \right) , \hspace{0.5in}
    [\mathcal{Q}]=\left(\begin{array}{ccc}
  1, & 0, & -1 \\
0,& 0, & 0 \\
   -1, &0, &1 \\
    \end{array}
    \right),     \nonumber
\end{eqnarray}
\begin{eqnarray}
[\mathcal{R}]=\left(\begin{array}{ccc}
   0,& 0, & 0 \\
0,& 1, & -1\\
    0,& -1, & 1 \\
    \end{array}
    \right) , \hspace{0.5in}
    [\mathcal{S}]=\left(\begin{array}{ccc}
  0, & 1, & -1 \\
 -1, & 0, & 1 \\
    1, & -1, &0 \\
    \end{array}
    \right),   
\end{eqnarray}
where $[\mathcal{P}]$, $[\mathcal{Q}]$, $[\mathcal{R}]$ are symmetric, and $[\mathcal{S}]$ is
(proportional to) the unique antisymmetric matrix for which sums of its columns and rows all vanish.
Given $[W_{CKM}]$ as in Eq.~(\ref{appW}), we may use the weighted mean values to find
\begin{eqnarray}
p=-0.05063 \pm 0.00010,   \hspace{0.2in} q=(-4.5 \pm 0.3)\times 10^{-5}, \nonumber \\
r= (-1.66 \pm 0.06) \times 10^{-3}, \hspace{0.2in}  s= (3.2 \pm 0.3)\times 10^{-5}.
\end{eqnarray}


\section{Some explicit formulas}

Because of the central role played by $\Lambda_{\gamma k}$,
it is useful to list explicitly $\Lambda_{\gamma k}$ in terms of the 
variables $(b,c,d,e)$ in the general parametrization given in Eq.~(\ref{bcde}).
We find
\begin{eqnarray}\label{Lambda-bcde}
2\Lambda_{e1}&=&-\frac{1}{3}-\frac{b}{2}+\frac{5}{12} c-\frac{1}{2} bc-\frac{1}{4} c^{2}+d(d+2e) \nonumber \\
2\Lambda_{e2}&=&-\frac{1}{6}+\frac{b}{2}+\frac{7}{12} c+\frac{1}{2} bc-\frac{1}{4} c^{2}+d(d-2e) \nonumber \\
2\Lambda_{e3}&=&\frac{1}{9}-\frac{3}{4} c-\frac{b}{6}-\frac{b^{2}}{2}+\frac{c^{2}}{8}+(\frac{d}{2}+e)(-d+2e) \nonumber \\
2\Lambda_{\mu 1}&=&-\frac{c}{3}+\frac{d}{6}-e+(bd+ce)+c(b+\frac{c}{2}+d) \nonumber \\
2\Lambda_{\mu 2}&=&-\frac{2}{3} c-\frac{d}{6}+e-(bd+ce)+c(-b+\frac{c}{2}+d) \nonumber \\
2\Lambda_{\mu 3}&=&-\frac{2}{9}+\frac{b}{3}+\frac{c}{2}-\frac{d}{2}+\frac{e}{3}+b^{2}-\frac{c^{2}}{4}+2be-\frac{cd}{2} \nonumber \\
2\Lambda_{\tau 1}&=&-\frac{c}{3}-\frac{d}{6}+e-(bd+ce)+c(b+\frac{c}{2}-d) \nonumber \\
2\Lambda_{\tau 2}&=&-\frac{2}{3} c+\frac{d}{6}-e+(bd+ce)+c(-b+\frac{c}{2}-d) \nonumber \\
2\Lambda_{\tau 3}&=&-\frac{2}{9}+\frac{b}{3}+\frac{c}{2}+\frac{d}{2}-\frac{e}{3}+b^{2}-\frac{c^{2}}{4}+\frac{1}{2} cd-2be. 
\end{eqnarray}

For vacuum mixing, the values $(b_{0},c_{0},d_{0},e_{0})$ are all small and we may ignore quadratic terms,
resulting in the matrix $[\Lambda(0)]$ given by Eq.~(\ref{LambdaA0}).

For the parametrization of $[W(0)]$ in vacuum, we expect to see a slight difference 
between NO and IO for $W_{\mu i}$ and $W_{\tau i}$ due to the uncertain measurements 
of $\sin^{2}\theta_{23}$ in SP,
as indicated by the estimations in Eqs.~(\ref{Winitial-NO}) and (\ref{Winitial-IO}). 
In addition to $[W(0)]_{N}$ under NO, with the parameters $(b_{0},c_{0},d_{0},e_{0})$ as in Eq.~(\ref{para}), 
here we propose to parametrize $[W(0)]$ under IO with $(b'_{0},c'_{0},d'_{0},e'_{0})$ as
\begin{equation}\label{B1}
[W(0)]_{I} = \left(\begin{array}{ccc}
  \smallskip
  \smallskip
   \frac{2}{3}+b'_{0}-\frac{c'_{0}}{2}, & \frac{1}{3}-b'_{0}-\frac{c'_{0}}{2}, & c'_{0} \\ 
   \smallskip
   \smallskip
   \frac{1}{6}-\frac{b'_{0}}{2}+\frac{c'_{0}}{4}-\frac{d'_{0}}{2}+e'_{0}, & \frac{1}{3}+\frac{b'_{0}}{2}+\frac{c'_{0}}{4}-\frac{d'_{0}}{2}-e'_{0}, &\frac{1}{2}+d'_{0}-\frac{c'_{0}}{2} \\
   \smallskip
   \smallskip
   \frac{1}{6}-\frac{b'_{0}}{2}+\frac{c'_{0}}{4}+\frac{d'_{0}}{2}-e'_{0}, & \frac{1}{3}+\frac{b'_{0}}{2}+\frac{c'_{0}}{4}+\frac{d'_{0}}{2}+e'_{0}, & \frac{1}{2}-d'_{0}-\frac{c'_{0}}{2} \\
    \end{array}
    \right),
\end{equation}
where $-d'_{0} \cong d_{0}$.
The $\Lambda$ matrix is then given by Eq.~(\ref{Lambda-bcde}) with the replacement of
$(b,c,d,e)$ by $(b'_{0},c'_{0},d'_{0},e'_{0})$.


Taking into account the possible distinction of $W(A_{i})$ between NO (Eq.~(\ref{WAi}))
and IO for $A \sim A_{i}$, as can be seen in
Fig. 1, we may further parametrize $W(A_{i})$ under IO as
\begin{equation}\label{WAiIO}
[W(A_{i})]_{I} \cong \left(\begin{array}{ccc}
   0, & 1-c'_{0}, & c'_{0} \\
   \smallskip
   \frac{1}{2}-d'_{0}, & c'_{0}/2, & \frac{1}{2}-(c'_{0}/2)+d'_{0} \\
   \smallskip
    \frac{1}{2}+d'_{0}, & c'_{0}/2, & \frac{1}{2}-(c'_{0}/2)-d'_{0} \\
    \end{array}
    \right),
  \end{equation}
from which the expression of $\Lambda$ under IO is given by Eq.~(\ref{LambdaAi}), after
changing $(b_{0},c_{0},d_{0},e_{0})$ to  $(b'_{0},c'_{0},d'_{0},e'_{0})$

For a very dense media, the matrix $[W(A_{d})]_{I}$ can be approximated by 
\begin{equation}\label{WAdIO}
[W(A_{d})]_{I} 
 \approx
 \left(\begin{array}{ccc}
   0, &1, & 0 \\
   \smallskip
  \frac{1}{2}-d'_{0}, &0, & \frac{1}{2}+d'_{0} \\
   \smallskip
    \frac{1}{2}+d'_{0}, & 0, & \frac{1}{2}-d'_{0} \\
    \end{array}
    \right),
  \end{equation}  
which yields the only nonvanishing element $\Lambda_{e2} \approx -(1/4)+d'^{2}_{0}$ under IO,
while from Eq.~(\ref{WAd}), the only nonvanishing element under NO is given by $\Lambda_{e3} \approx (-1/4)+d^{2}_{0}$.



\begin{thebibliography}{99}


\bibitem{Forero:2014bxa} 
  D.~V.~Forero, M.~Tortola and J.~W.~F.~Valle,
  Phys.\ Rev.\ D {\bf 90}, 093006 (2014).

\bibitem{Gonzalez-Garcia:2015qrr} 
  M.~C.~Gonzalez-Garcia, M.~Maltoni and T.~Schwetz,
  Nucl.\ Phys.\ B {\bf908}, 199 (2016).
  
\bibitem{Capozzi:2016rtj} 
  F.~Capozzi, E.~Lisi, A.~Marrone, D.~Montanino and A.~Palazzo,
  Nucl.\ Phys.\ B {\bf908}, 218 (2016).

\bibitem{Schechter:1980gk} 
  J.~Schechter and J.~W.~F.~Valle,
  Phys.\ Rev.\ D {\bf 23}, 1666 (1981).

\bibitem{Chiu:2015ega} 
  S.~H.~Chiu and T.~K.~Kuo,
  Phys.\ Lett.\ B {\bf 760}, 544 (2016).
  
\bibitem{Chiu:2016qra} 
  S.~H.~Chiu and T.~K.~Kuo,
  Phys.\ Rev.\ D {\bf 93}, 093006 (2016).
  


\bibitem{Kuo:2005pf} 
  T.~K.~Kuo and T.~H.~Lee,
  Phys.\ Rev.\ D {\bf 71}, 093011 (2005).

 \bibitem{Jarlskog:1985ht} 
  C.~Jarlskog,
  Phys.\ Rev.\ Lett.\  {\bf 55}, 1039 (1985). 

\bibitem{Chiu:2011rj} 
  S.~H.~Chiu and T.~K.~Kuo,
  Phys.\ Rev.\ D {\bf 84}, 013001 (2011).

\bibitem{Chiu:2012uc} 
  S.~H.~Chiu and T.~K.~Kuo,
  Eur.\ Phys.\ J.\ C {\bf 73}, 2579 (2013).


\bibitem{PDG}
C. Patrignani {\it et. al.}, (Particle Data Group), Chin. Phys. C 40, 100001 (2016)


\bibitem{Wolfenstein:1977ue} 
  L.~Wolfenstein,
  Phys.\ Rev.\ D {\bf 17}, 2369 (1978).
  
\bibitem{Mikheev:1986gs} 
  S.~P.~Mikheev and A.~Y.~Smirnov,
  Sov.\ J.\ Nucl.\ Phys.\  {\bf 42}, 913 (1985)
  [Yad.\ Fiz.\  {\bf 42}, 1441 (1985)].


\bibitem{Chiu:2010da} 
  S.~H.~Chiu, T.~K.~Kuo and L.~X.~Liu,
  Phys.\ Lett.\ B {\bf 687}, 184 (2010).


\bibitem{Naumov:1993vz} 
  V.~A.~Naumov,
  Phys.\ Lett.\ B {\bf 323}, 351 (1994).


\bibitem{Harrison:1999df} 
  P.~F.~Harrison and W.~G.~Scott,
  Phys.\ Lett.\ B {\bf 476}, 349 (2000).
  

\bibitem{Toshev:1991ku} 
  S.~Toshev,
  Mod.\ Phys.\ Lett.\ A {\bf 6}, 455 (1991).
\bibitem{LBL-1}
C. H. Albright {\it et. al.},
arXiv:physics/0411123 

\bibitem{Bandyopadhyay:2007kx} 
  A.~Bandyopadhyay {\it et al.} [ISS Physics Working Group],
  Rep.\ Prog.\ Phys.\  {\bf 72}, 106201 (2009).


\bibitem{Kuo:1989qe} 
  T.~K.~Kuo and J.~T.~Pantaleone,
  Rev.\ Mod.\ Phys.\  {\bf 61}, 937 (1989).

\bibitem{Dighe:1999bi} 
  A.~S.~Dighe and A.~Y.~Smirnov,
  Phys.\ Rev.\ D {\bf 62}, 033007 (2000).




	
\end{thebibliography}
\end{document}